\documentclass[preprint,11pt]{article}

\usepackage{graphicx,epsf,amsbsy,amsmath,amssymb}
\usepackage[latin1]{inputenc}

\usepackage{graphicx}

\begin{document}

\title{Poincar\'{e}'s forgotten conferences on wireless telegraphy}

\author{Jean-Marc Ginoux\footnote{Institut de Math\'{e}matiques de Jussieu, Universit\'e Pierre \& Marie Curie, Paris VI, UMR 7586, France, jmginoux@orange.fr, http://ginoux.univ-tln.fr} and Loic Petitgirard \footnote{Mus\'{e}e des arts et m\'{e}tiers, Conservatoire national des arts et m\'{e}tiers, France, loic.petitgirard@cnam.fr}}

\maketitle

\begin{abstract}
At the beginning of the twentieth century while Henri Poincar\'{e} (1854-1912) was already deeply involved in the developments of wireless telegraphy, he was invited, in 1908, to give a series of lectures at the \'{E}cole Sup\'{e}rieure des Postes et T\'{e}l\'{e}graphes (today Sup'T\'{e}lecom). In the last part of his presentation he established that the necessary condition for the existence of a stable regime of maintained oscillations in a device of radio engineering completely analogous to the triode: the singing arc, is the presence in the phase plane of stable limit cycle. The aim of this work is to prove that the correspondence highlighted by Andronov between the periodic solution of a non-linear second order differential equation and Poincar\'{e}'s concept of limit cycle has been carried out by Poincar\'{e} himself, twenty years before in these forgotten conferences of 1908.

\end{abstract}

\date{{\bf Keywords}: maintained oscillations, wireless telegraphy, limit cycles, stability, singing arc}

\section{Introduction}

The famous correspondence established by the Russian mathematician Aleksandr Andronov (1901-1952) in a note published in the \textit{Comptes Rendus} of the French Academy of Sciences in 1929 was until now considered by scientists and historians of science as a key moment in the development of the theory of nonlinear oscillations. One of the first to point out the importance of this result was Leonid Mandel'shtam (1879-1944), the Ph-D advisor of Andronov, during the sixth General Assembly of the Union Radio-Scientifique Internationale (U.R.S.I.)\footnote{in English: International Union of Radio Science, See \cite{vanbladel}.}:

\begin{quote}
``The relationship between the work of Poincar\'{e}, improved by Birkhoff, and those of Lyapunov, and our physical problem was reported by one of us\footnote{\cite{Andro29}. However, it will be established in the third section that the first contribution of Andronov on this subject was originally published in August 1928.}.  Three things should be distinguished here. First the qualitative theory of differential equations developed by Poincar\'{e} proved very efficient for qualitative discussion of physical phenomena that occur in systems used by radio engineering. But neither the physician nor a fortiori the engineer can not be satisfied with a qualitative analysis. Another series of works of Poincar\'{e} provides a method that enables to analyze our problems quantitatively. Finally the work of Lyapunov can give a mathematical discussion of the questions of stability.'' \cite[p. 83]{Mandel}
\end{quote}

A few years later, Nicolas Minorsky (1885-1970) wrote in his ``Introduction to Non-Linear-Mechanics'':

\begin{quote}
``Andronow\footnote{\cite{Andro29}.} was first to suggest that periodic phenomena in non-linear and non-conservative systems can be described mathematically in terms of limit cycles which thus made it possible to establish a connection between these phenomena and the theory of Poincar\'{e} developed for entirely different purposes.'' \cite[p. 63]{Min}
\end{quote}

Since then, many scientists and historians of science have considered Andronov as the first to have emphasized a connection with Poincar\'{e}'s works\footnote{See also \cite{Pechenkin}}.

\begin{quote}
``Henceforth, by using, transposing, or extending Poincar\'{e}'s arsenal Andronov would endeavor to develop Mandel'shtam's program. Also, reaping Lyapounov's heritage, Andronov focused on the problem of stability. Combining Poincar\'{e}'s small-parameter method with Lyapounov's stability theory, he established a method for finding periodic solutions and studying their stability.''\\ \vphantom{N} \hspace{4.2cm} \cite[p. 286]{AubinDahan2002}
\end{quote}

As above mentioned, let's notice that the correspondence of Andronov does not only deal with the analogy between the shape of the periodic solution of a nonlinear second order differential equation and Poincar\'{e}'s concept of limit cycle. In fact, the result of Andronov is of much greater importance since it concerns the stability of the limit cycle. In other words, it states that the necessary condition for establishing a stable regime of maintained oscillations\footnote{Such maintained oscillations will be designed by Andronov \cite{Andro29} as self-oscillations or self-maintained oscillations.} in a system (a radio engineering device for example) is the existence, in the phase plane, of a stable limit cycle.

Generally, Andronov's result is associated with that of Balthazar Van der Pol (1889-1959) who is wrongly credited for having highlighted the existence of a limit cycle in an oscillating circuit comprising a triode\footnote{In a series of publications, Van der Pol studied maintained oscillations by a triode. He plotted the periodic solution of this system by means of graphical integration (isoclines) and found that it was shaped like a closed curve. Unfortunately, he did not realized that this closed curve was a limit cycle of Poincar\'{e} as it is easy to check it in his publication \cite{VdP1926}.}. Although the triode was invented in 1907, its use was widespread only after the first World War. But at this time, Poincar\'{e} had already died prematurely. So, the question that arises then is the following:

\begin{center}
What kind of device has been employed by Poincar\'{e}\\ to observe maintained oscillations?
\end{center}

Before the advent of the triode, a device was commonly used in wireless telegraphy: the singing arc. Completely analogous\footnote{It will be established ten years later by Paul Janet \cite{Janet} that both triode and singing arc are completely analogous and are thus modeled by the same equation.} to the triode the singing arc was used to generate electromagnetic waves (radio waves).\\

During the last two decades of his life, Poincar\'{e} had been involved in many research on the propagation of electromagnetic waves. In 1890, he wrote to Hertz to report a miscalculation in his famous experiments\footnote{See \cite{Whittaker}}. Three years later, he solved the telegraphists equation \cite{Poin1893}. The following year he published a book entitled: ``Oscillations \'{e}lectriques'' \cite{Poin1894} and in 1899 another one: ``La Th\'{e}orie de Maxwell et les oscillations hertziennes'' \cite{Poin1899}. This book, also published in English and German in 1904 and reprinted in French in 1907, has been considered as a reference. In Chapter XIII, page 79 Poincar\'{e} stated that the singing arc and the Hertz spark gap transmitter were also analogous except that oscillations are maintained in the first and damped in the second. Thus, from the early twentieth century until his death Poincar\'{e} continued his research on wireless telegraphy and on maintained waves and oscillations [Poincar\'{e}, 1901, 1902, 1903, 1904, 1907, 1908, 1909abcde, 1910abc, 1911,1912].

On July $4^{th}$, 1902 he became Professor of Theoretical Electricity at the \'{E}cole Sup\'{e}rieure des Postes et T\'{e}l\'{e}graphes (today Sup'T\'{e}lecom) in Paris where he taught until 1910. The director of this school, \'{E}douard Estauni\'{e} (1862-1942), also asked him to give a series of conferences every two years. In 1908, Poincar\'{e} chose as the subject: wireless telegraphy. The text of his lectures was first published weekly in the journal \textit{La Lumi\`{e}re \'{e}lectrique} \cite{Poin1908} before being edited as a book \cite{Poin1909d}.

In the fifth and last part of these lectures entitled: ``T\'{e}l\'{e}graphie dirig\'{e}e~: oscillations entretenues\footnote{``Directive telegraphy: maintained oscillations.''}'' Poincar\'{e} stated a necessary condition for the establishment of a stable regime of maintained oscillations in the singing arc. More precisely, he demonstrated the existence, in the phase plane, of a stable limit cycle.

This paper is organized as follows. In the second section the fifth part of the Poincar\'{e}'s conferences \cite{Poin1908} will be fully presented and analyzed. Then, it will be compared to Andronov's work of 1929 \cite{Andro29} presented in the third section and it will be shown in fourth section that Poincar\'{e} and Andronov results are completely identical. Thus, the reasons why this fundamental paper of Poincar\'{e} has remained in oblivion, for scientists and historians of science for more than one century, will be discussed in the last section.

\section{Poincar\'{e}'s forgotten conferences on wireless telegraphy}
\label{Poincar}

\begin{figure}[htbp]
\centerline{\includegraphics{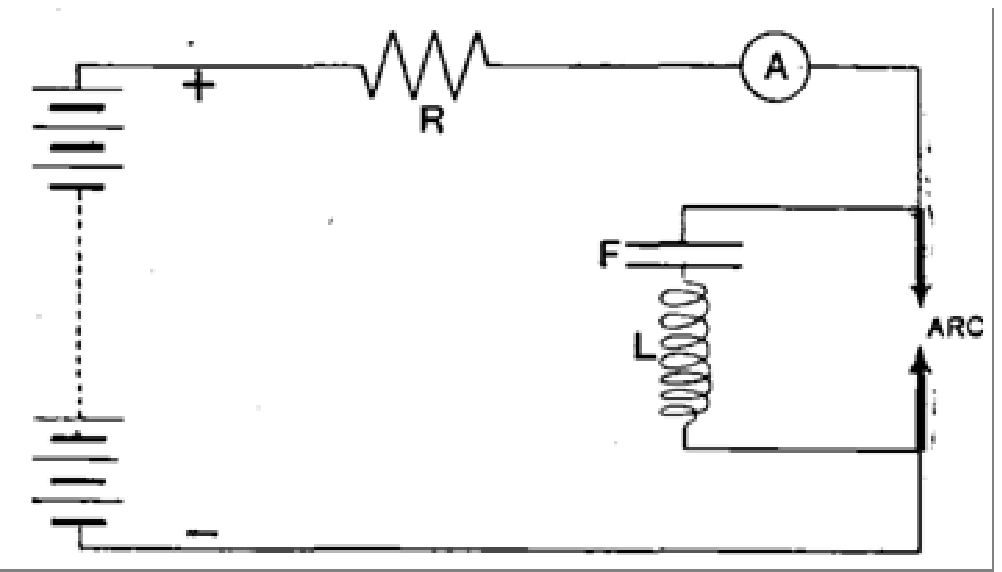}}
\caption{Singing arc circuit diagram \cite[p. 248]{Dud1900a}}
\label{fig1}
\end{figure}

At the end of the nineteenth a device, ancestor of the incandescent lamp, called electric arc was used for illumination of lighthouses and cities\footnote{The electrical arc (artificial in contrast to the flash of lightning) is associated with the electrical discharge produced between the ends of two electrodes (eg carbon), which also emits light. It is still used today in cinema projectors, plasma and thermal metallurgy in the ``arc welding'' or smelting (arc furnaces).}. It presented independently of its low light, a major drawback: the noise generated by the electrical discharge disturbed the residents. In London, the British physicist William Du Bois Duddell (1872-1917), was commissioned in 1899 by the English authorities to solve this problem. He had the idea of combining an oscillating circuit composed of inductor L and a capacitor of capacitance C (F in Fig. \ref{fig1}) electric arc to stop the rustling (see Fig. \ref{fig1}). After making such a device he called singing arc\footnote{For a brief history of the arc consult the work of Hertha Ayrton \cite[p. 19]{Ayrton1902}.}, Duddell \cite{Dud1900a,Dud1900b} then established that the musical sound\footnote{If its frequency is audible for human beings.} emitted by the arc corresponded to the period of oscillation circuit associated with it and expressed using the formula of Thomson \cite{Tho1853}.

In fact, Duddell had invented an oscillating circuit susceptible to produce sounds and more than that: electromagnetic waves. Thus, this apparatus will be used as emitter and receiver for the wireless telegraphy till the advent of the triode. By producing spark, the singing arc or Duddell generated electromagnetic waves highlighted by Hertz experiments \cite{Hertz}.

\pagebreak

\subsection{The singing arc equation}

\begin{figure}[htbp]
\centerline{\includegraphics{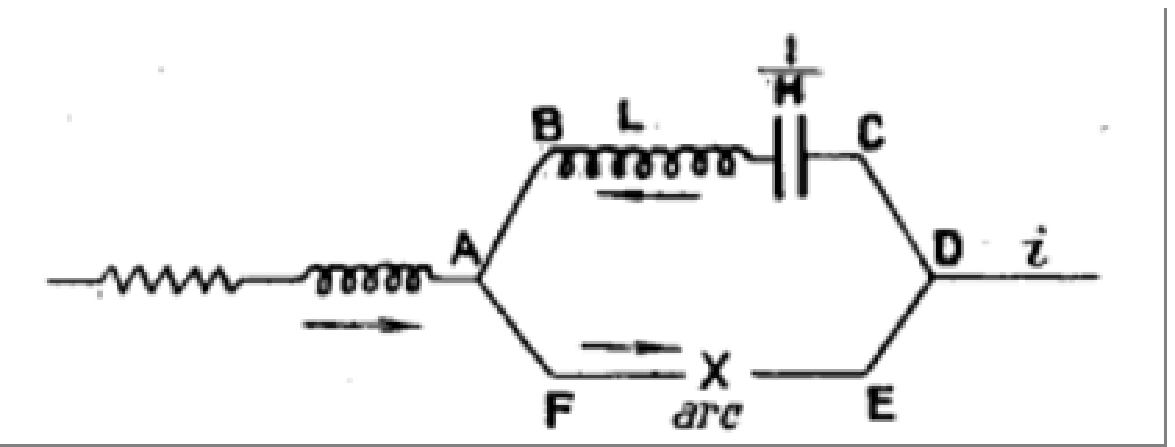}}
\caption{Maintained oscillations in the singing arc \cite[p. 390]{Poin1908}}
\label{fig2}
\end{figure}

In the last part of his lectures, Poincar\'{e} \cite[p. 390]{Poin1908} focused on the maintained oscillations in a singing arc circuit. The circuit diagram he studied (see Fig. \ref{fig2}) is completely identical to that of Duddell (see Fig. \ref{fig1}).

According to Poincar\'{e} \cite[p. 390]{Poin1908} this circuit consists of an Electro Motive Force (E.M.F.) of direct current E, a resistance R and a self, and in parallel, a singing arc and another self L and a capacitor. In order to provide the differential equation modeling the maintained oscillations he calls $x$ the capacitor charge and $i$ the current in the external circuit. Thus, the intensity in the branch (ABCD) comprising the capacitor of capacity $1/H$ may be written:

\[
x' = \frac{dx}{dt}
\]

The current intensity $i_{a}$ in the branch (AFED) comprising the singing may be written while using Kirchoff's law: $i_{a} = i + x'$. Then, Poincar\'{e} establishes the following second order nonlinear differential equation for the maintained oscillations in the singing arc:

\begin{equation}
  \label{Poin1}
Lx'' + \rho x' + \varphi \left( i + x' \right) + H x = 0
\end{equation}

He specifies that the term $\rho x'$ corresponds to the internal resistance of the self and various damping while the term $\varphi \left( i + x' \right)$ represents the E.M.F. of the arc which is related to the intensity by a function, unknown at that time. The main problem of equation (\ref{Poin1}) is that it depends on two variables $x$ and $i$. So, it is necessary for Poincar\'{e} to get rid of $i$. 

\newpage

By neglecting the external self and while equaling the tension in all branches of the circuit he finds that:

\begin{equation}
  \label{Poin2}
R i + \varphi \left( i + x' \right) = E
\end{equation}

He explains that if the function $\varphi$ was known, equation (\ref{Poin2}) would provide a relation between $i$ and $x'$ or between $i + x'$ and $x$ and then the variable $i$ could be eliminated in the equation (\ref{Poin1}). Thus, he makes the assumption\footnote{Probably based on the use of the \textit{Implicit Function Theorem}} that there exists a function relating $i$ and $x'$. Then, he directly replaces in equation (\ref{Poin1}) $\varphi \left( i + x' \right)$ by $\theta \left( x' \right)$ and writes:

\begin{equation}
  \label{Poin3}
Lx'' + \rho x' + \theta \left( x' \right) + H x = 0
\end{equation}

\subsection{Stability condition: \textit{Maintained oscillations and limit cycles}}

Then, Poincar\'{e} establishes, twenty years before Andronov \cite{Andro29}, that the stability of the periodic solution of equation (\ref{Poin3}) depends on the existence of a ``closed curve'', \textit{i.e.} a stable limit cycle in the phase plane. By using the variable changes he has introduced in his famous memoirs entitled: ``Sur les courbes d'\'{e}finis par une \'{e}quation diff\'{e}rentielle'' \cite[p. 168]{Poin1886} he sets:

\[
x' = \frac{dx}{dt} = y \quad \mbox{;} \quad dt = \frac{dx}{dy} = y \quad \mbox{;} \quad x'' = \frac{dy}{dt} = \frac{ydy}{dx}
\]

Thus, equation (\ref{Poin3}) becomes:

\begin{equation}
  \label{Poin4}
Ly\frac{dy}{dx}  + \rho y + \theta \left( y \right) + H x = 0
\end{equation}

Poincar\'{e} states then that:

\begin{center}
``Les oscillations entretenues correspondent aux\\ courbes ferm\'{e}es, s'il y en a.\footnote{``Maintained oscillations correspond to closed curves if there exist any.''}\cite{Poin1908}'' \end{center}

\vspace{0.1in}

and he gives the following representation for the solution of equation (\ref{Poin4}):

\begin{figure}[htbp]
\centerline{\includegraphics{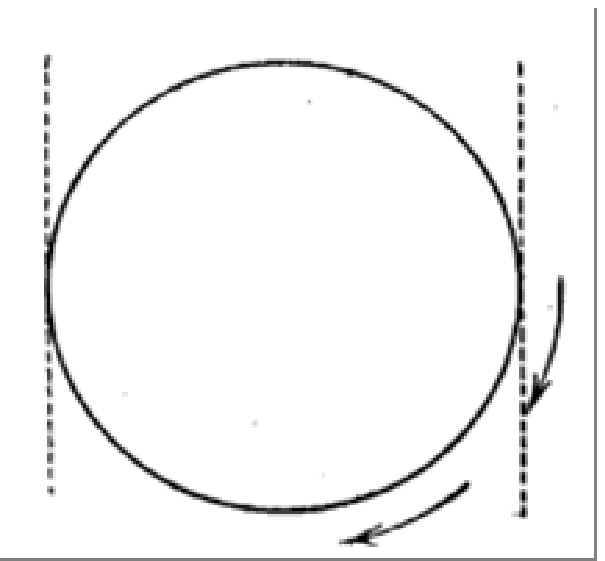}}
\caption{Closed curve solution of equation (\ref{Poin4}) \cite[p. 390]{Poin1908}}
\label{fig3}
\end{figure}

Let's notice that this closed curve is only a ``metaphor'' of the solution since Poincar\'{e} do not use any graphical integration method such as ``isoclines\footnote{Van der Pol \cite{VdP1926} is wrongly credited for the invention of the ``method of isoclines''. In fact, this method has been introduced in 1887 by a Belgian engineer named Junius Massau \cite[p. 501]{Massau}.}''. Moreover, the main purpose of this representation is to specify the sense of rotation of the trajectory curve which is a preliminary necessary condition to the establishment of the following proof involving the \textit{Green-Ostrogradsky} theorem.\\

Then, Poincar\'{e} explains that if $y = 0$ then $dy/dx$ is infinite and so, the curve admits vertical tangents. Moreover, if $x$ decreases $x'$, \textit{i.e.} $y$ is negative. He concludes that the trajectory curves turns in the direction indicated by the arrow (see Fig. \ref{fig3}).\\

\pagebreak

Poincar\'{e} writes:

\begin{quote}
``Condition de stabilit\'{e}. - Consid\'{e}rons donc une autre courbe non ferm\'{e}e satisfaisant \`{a} l'\'{e}quation diff\'{e}rentielle, ce sera une sorte de spirale se rapprochant ind\'{e}finiment de la courbe ferm\'{e}e. Si la courbe ferm\'{e}e repr\'{e}sente un r\'{e}gime stable, en d\'{e}crivant la spirale dans le sens de la fl\`{e}che on doit \^{e}tre ramen\'{e} sur la courbe ferm\'{e}e, et c'est \`{a} cette seule condition que la courbe ferm\'{e}e repr\'{e}sentera un r\'{e}gime stable d'ondes entretenues et donnera lieu à la solution du probl\`{e}me.\footnote{``Stability condition. - Let's consider another non-closed curve satisfying the differential equation, it will be a kind of spiral curve approaching indefinitely near the closed curve (so called limit cycle). If the closed curve represents a stable regime, by following the spiral in the direction of the arrow one should be brought back to the closed curve, and provided that this condition is fulfilled the closed curve will represent a stable regime of maintained waves and will give rise to a solution of this problem.''}'' \cite[p. 391]{Poin1908}
\end{quote}

In the \textit{Notice sur les Travaux scientifiques d'Henri Poincar\'{e}} he wrote in 1886, he defines the concept of limit cycle:

\begin{quote}
``J'appelle ainsi les courbes ferm\'{e}es qui satisfont à notre \'{e}quation diff\'{e}rentielle et dont les autres courbes d\'{e}finies par la m\^{e}me \'{e}quation se rapprochent asymptotiquement sans jamais les atteindre.\footnote{``I call thus closed curves that satisfy our differential equation and whose other curves defined by the same equation are approaching asymptotically without never reaching them.''}'' \cite[p. 30]{Poin1886n}
\end{quote}

By comparing both definitions it clearly appears that the ``closed curve'' which represents a stable regime of maintained oscillations is nothing else but a limit cycle as Poincar\'{e} has defined it in his own works. But this, first ``giant step'' is not sufficient to prove the stability of the oscillating regime. Poincar\'{e} has to demonstrate now that the periodic solution of equation (\ref{Poin3}) (the ``closed curve'') corresponds to a stable limit cycle.

\subsection{Possibility condition of the problem: \textit{stability of limit cycles}}

In the following part of his lectures, Poincar\'{e} gives what he calls  a ``condition de possibilit\'{e} du probl\`{e}me''. In fact, he establishes a condition of stability of the periodic solution of equation (\ref{Poin3}), \textit{i.e.} a condition of stability of the limit cycle under the form of inequality.\\

After multiplying equation (\ref{Poin4}) by $x'dt$ Poincar\'{e} integrates it over one period while taking into account that the first and fourth term are vanishing since they correspond to the conservative part of this nonlinear equation\footnote{It is easy to show that: $\int {Lydy} + \int {Hxdx} = \frac{1}{2}Ly^2 + \frac{1}{2}Hx^2 = 0$ }. He finds:

\begin{equation}
  \label{Poin5}
\rho \int {{x}'^2dt} +\int {\theta \left( {{x}'} \right){x}'dt} = 0
\end{equation}

Then, he explains that since the first term is quadratic, the second one must be negative in order to satisfy this equality. So, he stated that the oscillating regime is stable iff:

\begin{equation}
  \label{Poin6}
\int {\theta \left( {{x}'} \right){x}'dt} < 0
\end{equation}

It will be shown in the next section that this inequality is completely identical to the one Andronov \cite{Andro29} will state twenty years later.

\section{Andronov's works on self-oscillations}
\label{Andro}

In 1920, Aleksandr Aleksandrovich Andronov (1901-1952) entered the Electrical Engineering Department of the Technical High-School of Moscow where a radio
engineering specialization was proposed. Five years later, he obtained a diploma in Theoretical Physics (Master Degree) at the university of Moscow. Then, he started a Ph-D with Leonid Isaakovich Mandel'shtam (1879-1944). This charismatic figure which is at the origin of the concept of ``nonlinear thinking\footnote{See Rytov \cite[p. 172]{Rytov}.}'' has deeply influenced the young Andronov. In fact, the correspondence he established in the famous note at the \textit{Comptes Rendus} was preceded by a short presentation of his Ph-D works\footnote{To our knowledge Andronov's Ph-D thesis has not been located nor referenced till now even in Andronov's works.} at the sixth congress of Russian Physicists at Moscow between the $5^{th}$ and $16^{th}$ August 1928 \cite{Andro28}. In this work Andronov gives the foundations of what will become the theory of nonlinear oscillations.

\begin{quote}
``However, any sufficiently rigorous general theory for such oscillations does not exist nowadays. Meanwhile, there is an adequate mathematical model or schema, created without any connection with the theory of oscillations, which allows a common view of all these processes to the case of one degree of freedom. This concept is the ``theory of limit cycles'' of Poincar\'{e}.'' \cite[p. 23]{Andro28}
\end{quote}

In his conclusion, which should be compared to that of Poincar\'{e} \cite[p. 391]{Poin1908} (See above p. 6), Andronov introduced his famous neologism\footnote{According to Pechenkin, Andronov has invented this terminology ``by combining the Greek word ``$a\nu to$'' (``auto'') with Russian word ``kolebania'' (``oscillations'') \cite[p. 288]{Pechenkin}. In fact, it seems that Andronov was inspired by the reading of Heinrich Barkhausen (1881-1956) who used in his Ph-D dissertation in 1907 the German expression ``selbst Schwingungen'' (self-oscillations). See \cite[p. 59]{Barkhausen1907} and also \cite[p. 561]{Andro29} }:

\begin{quote}
``The stable motions existing in devices capable of self-oscillations must always correspond to limit cycles.''\cite[p. 24]{Andro28}
\end{quote}

On Monday, October $14^{th}$, 1929 the French mathematician Jacques Hadamard (1865-1963) presented to the Academy of Sciences of Paris a note from Alexander Andronov. The fact that Hadamard had presented this work is not really surprising since on the one hand he was responsible for mathematical analysis section at the Academy of Sciences and on the other hand he was also correspondent of the Russian Academy of Sciences since 1922 and a foreign member of the Academy of Sciences of the USSR since 1929\footnote{See \cite{Mazya}.}. In this work, Andronov considers first many examples of non-conservative systems such as the problem of Cepheids for P.D.E., the Froude pendulum and the triode oscillator for nonlinear O.D.E.

\begin{quote}
``Citons, pour le cas des \'{e}quations aux d\'{e}riv\'{e}es partielles, le probl\`{e}me d\'{e}j\`{a} ancien de la corde vibrante excit\'{e}e par un archet ainsi que le probl\`{e}me des C\'{e}ph\'{e}ides, tel que le traite Eddington $(^1)$; pour celui des \'{e}quations diff\'{e}rentielles ordinaires, en m\'{e}canique le pendule de Froude $(^2)$, en physique l'oscillateur \`{a} triode $(^3)$, en chimie les r\'{e}actions p\'{e}riodiques $(^4)$; des probl\`{e}mes similaires se posent en biologie $(^5)$.''\\

\small
(1) EDDINGTON, The internal constitution of stars, p. 200 (Cambridge, 1926).\\
(2) Lord RAYLEIGH, The theory of sound, London 1, 1894, p. 212.\\
(3) Voir par exemple VAN DER POL, Phil. Mag., 7 s\'{e}rie, 2, 1926, p. 978.
(4) Voir par exemple KREMANN, Die periodischen Erscheinungen in der Chemie, p. 124 (Stuttgart, 1913).\\
(5) LOTKA, Elements of physical biology, p. 88 (Baltimore, 1925). Voir aussi les r\'{e}centes recherches de M. Volterra.
\normalsize
\end{quote}

\begin{flushright}
\cite[p. 560]{Andro29}
\end{flushright}

\subsection{Self-oscillations and limit cycles}

Then, Andronov explains that such systems he calls ``self-oscillators`` can be represented in the phase plane by two simultaneous differential equations:

\begin{equation}
  \label{Andro1}
\frac{dx}{dt} = P\left( x, y \right) \quad \mbox{;} \quad \frac{dy}{dt} = Q\left( x, y \right)
\end{equation}

and he adds that:

\begin{quote}

``It may easily be shown that, to periodic motions satisfying these conditions, there correspond, in the $xy$ plane, isolated closed curves, approached in spiral fashion by neighboring solutions from the interior or the exterior (for increasing $t$). As a result, self-oscillations arising in systems characterized by equations of type (\ref{Andro1}) correspond mathematically to stable Poincar\'{e} limit cycles $(^3)$.\\

\small
$(^3)$ POINCAR\'{E}, OEuvres, I, p. 53 (Paris, 1928)
\normalsize

\end{quote}

\begin{flushright}
\cite[p. 560]{Andro29}
\end{flushright}

It is important to notice, on the one hand, that due to the imposed format of the \textit{Comptes Rendus}, Andronov does not provide any demonstration. He just claims that the periodic solution of a non-linear second order differential equation defined by (\ref{Andro1}) ``corresponds'' to stable Poincar\'{e} limit cycles. On the other hand, it is interesting to compare this sentence with that of Poincar\'{e} (See above p. 7). Then, it clearly appears that Andronov has stated the same correspondence as Poincar\'{e} twenty years after him. Nevertheless, it seems that Andronov may not have read Poincar\'{e}'s article since at that time even if the first volume of his complete works had been already published it didn't contained this paper.

\subsection{Stability condition of limit cycles}

The next step for Andronov is to show that the periodic solution, \textit{i.e.} the limit cycle is stable. To this purpose he considers the following system, where $\mu$ is a real parameter, as an example:

\begin{equation}
  \label{Andro2}
\frac{dx}{dt} = y + \mu f\left( x, y; \mu \right) \quad \mbox{;} \quad \frac{dy}{dt} = -x + \mu g\left( x, y; \mu \right)
\end{equation}

He explains that for $\mu = 0$ the solution of this system is: $x = R \cos\left( t \right)$, $y = R \sin\left( t \right)$ as it is obvious to check. This enables him to introduce an ``unusual\footnote{Unusual since it corresponds to a clockwise rotation and not to the classical counter clockwise trigonometric rotation. But, it corresponds exactly with the rotation direction of the trajectory curve such as Poincar\'{e} has established it. See Fig. \ref{fig3} p. 5.}'' variables changes in polar coordinates. Then, by using Poincar\'{e}'s methods \cite[tome I, p. 89]{PoinMNMC} he states that for sufficiently small $\mu \neq 0$, the $xy$ plane contains only isolated closed curves, near to circles with radii defined by the equation:

\begin{equation}
  \label{Andro3}
\int_0^{2\pi} \left[ f\left(R \cos \xi, - R \sin \xi ; 0  \right) \cos \xi - g\left(R \cos \xi, - R \sin \xi ; 0  \right) \sin \xi \right] d\xi = 0
\end{equation}

Andronov provides a stability condition for the steady-state motion, \textit{i.e.} for the limit cycle:

\begin{equation}
  \label{Andro4}
\int_0^{2\pi} \left[ f_x \left(R \cos \xi, - R \sin \xi ; 0  \right) \cos \xi + g_y \left(R \cos \xi, - R \sin \xi ; 0  \right) \sin \xi \right] d\xi < 0
\end{equation}

In fact, this condition is based on the use of \textit{characteristic exponents} introduced by Poincar\'{e} in his so-called \textit{New Methods on Celestial Mechanics} \cite[tome I, p. 161]{PoinMNMC} and after by Lyapounov in his famous textbook \textit{General Problem of Stability of the Motion} \cite{Lyapounov}. That's the reason why Andronov will call later the stability condition (\ref{Andro4}): stability in the sense of Lyapounov or Lyapounov stability. It will be stated in the next section that both stability condition of Poincar\'{e} (\ref{Poin6}) and of Andronov (\ref{Andro4}) are totally identical.

\section{Poincar\'{e} stability versus Lyapounov stability}

In order to establish a comparison between Poincar\'{e}'s results and that of Andronov it is necessary to transform Eq. (\ref{Poin4}) into a dimensionless system. This can be easily done by using this variables changes: $x \rightarrow \sqrt{L/H}$, $t \rightarrow \mu t$ and while posing: $\mu = 1/{\sqrt{LH}}$. Then, starting from Eq. (\ref{Poin4}) and by neglecting the resistance $\rho$ of the self we have:

\begin{equation}
  \label{Poin7}
  \left\{
    \begin{array}{l}
      \dfrac{dx}{dt} = y \vspace{4pt} \\
      \dfrac{dy}{dt} = -x - \mu \theta \left( t \right)
    \end{array}
  \right.
\end{equation}

By comparing with the system of Andronov (\ref{Andro2}) we find that: $f\left( x, y; \mu \right) = 0$ and $g\left( x, y; \mu \right) = - \theta \left( y \right)$. Moreover, the stability condition (\ref{Andro4}) is only the rough idea of a theorem which will be formalized later by Pontryagin \cite{Pontryagin}. This theorem involves the Green's formula (\cite[p. 100]{Pontryagin}):

\begin{equation}
  \label{Green}
\int_{\Gamma} f \left(x, y  \right) dy - g \left(x, y  \right) dx = \iint_{S} \left( f'_{x} \left(x, y  \right)  + g'_{y} \left(x, y  \right) \right) dxdy
\end{equation}

where $\Gamma$ and $S$ design respectively a closed path and a surface. Let's notice that this theorem may be only stated provided that the sense of rotation on the closed path (curve) has been previously defined or chosen. That's the reason why Poincar\'{e} has accurately specified it (see above p. 5). Then, by using Cartesian coordinates system it may be shown that the stability condition of Andronov (\ref{Andro4}) reads:

\begin{equation}
  \label{Poin8}
\int_{\Gamma} f \left(x, y; \mu  \right) \frac{dy}{dt} - g \left(x, y; \mu  \right) \frac{dx}{dt} < 0
\end{equation}

Finally, by replacing in Eq. (\ref{Poin8}) $f\left( x, y; \mu \right) = 0$ and $g\left( x, y; \mu \right) = - \theta \left( y \right)$ and taking into account Poincar\'{e}'s variables changes (see above p. 5): $x' = dx/dt = y$ we have:

\begin{equation}
  \label{Poin9}
\int_{\Gamma} \theta \left( t \right) x' dt < 0
\end{equation}

This condition (\ref{Poin9}) exactly transcribes the fact that the characteristic exponent or Lyapounov exponent is negative. So, the identity between both Poincar\'{e} and Andronov stability conditions is thus stated. Then, it appears that Poincar\'{e} had not only established a correspondence between maintained oscillations and the existence of a limit cycle but he had also proved the stability of this limit cycle through a condition that Andronov will find again (independently) two decades later.

\pagebreak

\section{Discussion}
\label{conc}

In this paper it has been proved that Poincar\'{e} in these ``forgotten'' conferences has established two correspondences between technical problems of oscillations coming from  wireless telegraphy and his own works. As well as Andronov in his note at the \textit{Comptes Rendus}. Indeed, both of them have used, on the one hand, the concept of limit cycle that Poincar\'{e} had introduced in his famous memoirs and, on the other hand, the concept of characteristic exponents he had developed for Celestial Mechanics (especially for periodic orbits) in his so-called New Methods.\\

But, while the former only represents a minor step towards the theory of nonlinear oscillations, because if the limit cycle is unstable no maintained oscillations can be observed, the later is of fundamental importance. It is very surprising to notice that many historians of science have only focussed on the former weakening thus the impact of this result. Moreover, it is difficult to explain why these conferences have been completely "forgotten" by both scientists and historians over more than a century.\\

Many hypotheses are to be considered. The main reason is probably that these conferences have never been published in Poincar\'{e}'s complete works, only in the journal \textit{La Lumi\`{e}re \'{e}lectrique} (which disappeared in 1916) and in a textbook. Moreover, they clearly tackle technological problems that are the concern of engineers rather than mathematicians. Papers referring specifically to these conferences address the question of diffraction of radio waves, not maintained oscillations.\\

No reference to these conferences has been found until today in the technological neither mathematical literature.
But other hypotheses must be stated. First it may be reminded that Poincar\'{e} studies the singing arc circuit and not the triode circuit. But, after 1920 the singing arc is considered as completely obsolete by engineers, maybe explaining partly that nobody cares about the result of 1908. Except the fact that both singing arc and triode are analogous devices and are so modeled by the same equations, but was this known largely in the 1920s?\\

Second is the fact that the conferences aimed at presenting the solution of a very ``difficult'' problem in 1908 to students in telegraphy engineering: the public did not have a high mathematical background, except for the curious who may have attended it. Considering also that during the war, which started in 1914, most of those students may have been killed and the memory of this work may have disappeared in the trenches.\\

For now, many questions are unresolved. For example: why Poincar\'{e} did not use the terminology limit cycle while he gives a very accurate definition of the closed curve towards which any non-closed curve tends? Is it why the audience was supposed to be engineers without basic notions of qualitative theory of differential equations? The problem is that we ignore who was precisely that day in the audience, and have no idea who may have read the texts, who may get inspired with it (without citing it).\\

In any case, it remains clear that this work of 1908 represents the first application of Poincar\'{e} research (on what is called today dynamical system) in a technological problem, anticipating thus the development of the theory of nonlinear oscillations in the twentieth century.

\pagebreak

\end{document}